\documentclass[prd,amsmath,amssymb,longbibliography,superscriptaddress,twocolumn,nofootinbib,10pt]{revtex4}

\pdfoutput=1

\usepackage{graphicx}
\usepackage{dcolumn}
\usepackage{bm}
\usepackage{amssymb}
\usepackage{latexsym}
\usepackage{booktabs}
\usepackage{amsmath}
\usepackage{multirow}
\usepackage[colorlinks=true, linkcolor=blue, citecolor=blue]{hyperref}
\usepackage{xcolor}
\usepackage[utf8]{inputenc} 

\begin{document}

\title{Probing potential redshift-dependent systematics in the Hubble tension: Model-independent $H_0$ constraints from DESI R2}

\author{Tonghua Liu}
\affiliation{School of Physics and Optoelectronic, Yangtze University, Jingzhou 434023, China;}
\author{Shuo Cao}
\email{caoshuo@bnu.edu.cn}
\affiliation{School of Physics and Astronomy, Beijing Normal University, Beijing 100875, China;}
\affiliation{Institute for Frontiers in Astronomy and Astrophysics, Beijing Normal University, Beijing 102206, China;}
\author{Jieci Wang}
\email{jcwang@hunnu.edu.cn}
\affiliation{Department of Physics, and Collaborative Innovation Center for Quantum Effects and Applications, Hunan Normal University, Changsha 410081, China;}

\begin{abstract}
{We present a determination of the Hubble constant ($H_0$) using the latest observational data from multiple cosmological probes, providing an independent geometric calibration of the SN Ia distance scale. By combining baryon acoustic oscillation (BAO) measurements from the second data release of the Dark Energy Spectroscopic Instrument (DESI DR2), cosmic chronometer $H(z)$ data, and the Pantheon Plus Type Ia supernova (SN Ia) sample, we reconstruct the cosmic expansion history through Gaussian process regression without assuming a specific cosmological model. Our analysis fully incorporates the complete covariance structure and yields $H_0$ constraints at five distinct redshifts: $65.72 \pm 1.99$ (z=0.51), $67.78 \pm 1.75$ (z=0.706), $70.74 \pm 1.39$ (z=0.934), $71.04 \pm 1.93$ (z=1.321), and $68.37 \pm 3.95~\mathrm{km~s^{-1}~Mpc^{-1}}$ (z=1.484). The Bayesian combination of these measurements gives $\hat{H}_0 = 69.29 \pm 0.81~\mathrm{km~s^{-1}~Mpc^{-1}}$ with 1.2\% precision, which occupies an intermediate position between the Planck CMB result and the SH0ES local measurement. While we observe a non-monotonic pattern in $H_0$ values across redshifts, statistical tests show this apparent evolution is not significant (p = 0.208). Our approach delivers independent constraints at multiple redshifts, enabling investigation of potential redshift-dependent systematic effects in the Hubble tension. The results demonstrate that an independent geometric method yields an $H_0$ value consistent with the intermediate range of current measurements, providing a crucial cross-check of distance ladder determinations.}
\end{abstract}

\maketitle

\section{Introduction}
The Hubble constant ($H_0$), quantifying the present expansion rate of the Universe, stands as one of the most fundamental parameters in modern cosmology. Its precise determination not only sheds light on the age and evolution of the cosmos but also serves as a critical test for the validity of the standard cosmological model. However, a persistent and statistically significant discrepancy has emerged between early-Universe constraints from the cosmic microwave background (CMB) \cite{2020A&A...641A...6P}  and late-Universe measurements using Type Ia supernova (SN Ia) \cite{1998AJ....116.1009R,1999ApJ...517..565P}, presenting one of the most pressing challenges in contemporary cosmology.
The Planck collaboration's analysis of CMB data within the $\Lambda$CDM framework yields $H_0 = 67.4 \pm 0.5~\mathrm{km~s^{-1}~Mpc^{-1}}$ \cite{2020A&A...641A...6P}, while the SH0ES team's measurement using SN Ia calibrated with Cepheid variables gives $H_0 = 73.04 \pm 1.04~\mathrm{km~s^{-1}~Mpc^{-1}}$ \cite{2022ApJ...934L...7R}. This tension, now exceeding $5\sigma$ significance, has prompted extensive investigations into potential systematic uncertainties or new physics beyond the standard model. Various independent measurements, including those from the Megamaser Cosmology Project ($H_0 = 73.9 \pm 3.0~\mathrm{km~s^{-1}~Mpc^{-1}}$) \cite{2020ApJ...891L...1P}, strong lensing time delays from TDCOSMO collaboration ($H_0 = 67.4^{+4.1}_{3.2}~\mathrm{km~s^{-1}~Mpc^{-1}}$) \cite{2020A&A...643A.165B} and the tip of the red giant branch (TRGB) method ($H_0 = 69.6\pm 2.5~\mathrm{km~s^{-1}~Mpc^{-1}}$) \cite{2020ApJ...891...57F}, further complicate this picture, suggesting that the discrepancy may reflect either unaccounted systematic effects or fundamental limitations in our cosmological framework. More discussions on Hubble tension please see the references \citep{2021APh...13102605D,2022JHEAp..34...49A,2025SciBu..70..829L,2025arXiv250404417C,2024PDU....4401464O,2025JHEAp..4800405D,2022Galax..10...24D,2020PhRvD.102j3525K,2023EPJC...83..495L,2023PDU....3901160B,2022A&A...668A.135S,2022NewAR..9501659P,2019PhRvL.122v1301P,2019PhRvL.122f1105F,2021ApJ...912..150D,2020ApJ...895L..29L,2022ApJ...939...37L,2023ApJS..264...46L,2021PhRvD.103h1305B} and  references therein for a more comprehensive discussion.

Baryon acoustic oscillations (BAO) provide a powerful geometric probe of cosmic expansion. As relics of sound waves propagating in the primordial plasma before recombination, BAO features imprinted in the large-scale structure distribution offer a standard ruler for precision distance measurements. The characteristic scale of $\sim150$ Mpc, corresponding to the sound horizon at the drag epoch, remains imprinted in the clustering pattern of galaxies, providing a cosmic yardstick that can be measured at various redshifts.
The Dark Energy Spectroscopic Instrument (DESI), currently completing its 5-year survey, has emerged as a powerful tool to address these challenges. Its Data Release 1 (DR1) \cite{2025JCAP...02..021A} and Data Release 2 (DR2) \cite{2025arXiv250314738D} have delivered ground-breaking results from spectroscopic observations of over 40 million galaxies and quasars, providing unprecedented baryon acoustic oscillation (BAO) measurements across $0.1 < z < 4.2$. While DR1 achieved sub-percent precision on Hubble parameter and angular diameter distance, DR2 has further improved these constraints by incorporating additional sky coverage and improved redshift measurements. The recent DR2 from the DESI represents a substantial advancement in BAO measurements \cite{2025arXiv250314738D}. With expanded sky coverage (14,200 square degrees), improved sample sizes (emission-line galaxies increased by factor of 2.7, quasars by 1.7), and enhanced observational completeness (from 35.2\% to 53.7\% for emission-line galaxies), DESI DR2 provides unprecedented precision in BAO measurements across the redshift range $0.1<z<4.2$. These improvements yield 30\%-50\% better statistical constraints compared to previous data releases, enabling more robust cosmological parameter estimation.

Unlike traditional approaches that rely on assumptions about the sound horizon scale $r_s$ or specific dark energy models, our analysis employs a {cosmological model-independent} framework. By combining DESI DR2 BAO measurements with cosmic chronometer $H(z)$ data and Pantheon Plus SN Ia samples through Gaussian process regression (GPR), we reconstruct the cosmic expansion history {with minimal assumptions about the underlying cosmology}. We implement a complete statistical framework that properly accounts for all relevant uncertainties and correlations, including the covariance structure of BAO measurements, uncertainties in SN Ia, cosmic chronometer $H(z)$ measurements, and correlations introduced by the GPR reconstruction process. {Our methodology builds upon several fundamental assumptions: the validity of the distance duality relation, the reliability of stellar population models in cosmic chronometer measurements, and the appropriateness of GPR for reconstructing smooth expansion histories.}
Rather than providing a single $H_0$ measurement, our approach yields independent constraints at distinct redshifts, enabling investigation of potential redshift-dependent systematic effects that may contribute to the Hubble tension. We employ a Monte Carlo sampling procedure that propagates all sources of uncertainty through the complete analysis pipeline, from data reconstruction to final parameter estimation, ensuring robust error estimation and proper handling of correlations between different redshift measurements.

This paper is organized as follows: Section 2 describes the methodological framework and datasets employed. Section 3 presents our main results and analysis. Section 4 summarizes our conclusions.

\section{Hubble Constant Measurement Methodology}\label{sec:h0_methodology}
In any metric theory of gravity where photons follow null geodesics and photon number is conserved, the distance duality relation (DDR) holds for all redshifts $z$ \cite{1933PMag...15..761E}:
\begin{equation}
\eta(z) \equiv \frac{D_L(z)}{(1+z)^2 D_A(z)} = 1.
\label{eq:DDR}
\end{equation}
This identity is derived from fundamental geometric principles. Recent studies on testing the DDR indicate that it holds with extremely high precision and is unlikely to be violated \cite{2025ApJ...979....2Q,2020JCAP...06..036G,2025arXiv250711518K,2025ApJ...987...58W,2021EPJC...81..903L}.

Assuming the validity of the DDR (i.e., $\eta=1$), Eq. (\ref{eq:DDR}) can be rearranged to express the Hubble constant $H_0$. Starting from the definition:
\[
\frac{D_L(z)}{(1+z)^2 D_A(z)} = 1,
\]
we multiply both sides by $H_0$ and divide by $H(z)$ (assuming $H(z) \neq 0$), yielding:
\[
H_0 = \frac{1}{(1+z)^2} \cdot \frac{H_0 D_L(z)}{H(z) D_A(z)} \cdot H(z).
\]
This form suggests that $H_0$ can be determined through a combination of observational probes:
\begin{equation}
H_0 = \frac{1}{(1+z)^2} \cdot \frac{[H_0 D_L(z)]^{\rm SN}}{[H(z) D_A(z)]^{\rm BAO}} \cdot [H(z)]^{\rm CC}.
\label{eq2}
\end{equation}
The determination of $H_0$ via this relation relies on three independent observational probes:
1). The combination $[H(z) D_A(z)]^{\rm BAO}$ is obtained from BAO measurements, using both line-of-sight and transverse clustering scales, which does not require an external calibration of the sound horizon; 2). The Hubble parameter $[H(z)]^{\rm CC}$ is measured directly using cosmic chronometers (CC), which compare differential ages of passively evolving galaxies (e.g., in globular clusters) with their spectroscopic redshifts. This approach provides model-independent $H(z)$ estimates, though it depends on certain astrophysical modeling assumptions. 3). The quantity $[H_0 D_L]^{\rm SN}$ is inferred from Type Ia supernova (SN Ia) observations, which also do not rely on external calibration. {This methodology was first introduced by \citet{2023PhRvD.107b3520R}, and was subsequently advanced by \citet{2025ApJ...978L..33G} who incorporated the latest observational data from DESI DR1.}

Next, we will introduce these three types of data respectively.

\subsection{Unanchored Luminosity Distance from observation of  SN Ia }
Type Ia supernovae (SN Ia), as standardizable candles, serve as powerful cosmological probes. Their observations led to the discovery of the accelerating expansion of the Universe.  We adopt SN Ia data from Pantheon Plus data.
Our analysis incorporates the full Pantheon Plus  SN Ia sample \cite{2022ApJ...938..113S}, comprising 1,701 high-quality light curves from 1,550 spectroscopically confirmed  SN Ia  spanning an extensive redshift range ($0.01<z<2.26$). To minimize potential systematics from calibration uncertainties at low redshifts, we implement two key methodological choices: (1) we exclude the calibration subsample $(z < 0.01)$ that may be affected by peculiar velocity corrections and host galaxy contamination, and (2) we rigorously account for the full covariance matrix$\footnote{\url{https://github.com/PantheonPlusSH0ES/DataRelease}}$ that captures both statistical uncertainties and systematic correlations between supernova measurements. This conservative approach ensures our cosmological constraints remain robust against calibration-related biases while maintaining the statistical power of the full dataset. The remaining sample of 1,485 SN Ia after this selection provides a well-characterized Hubble diagram for precision cosmology.

The distance modulus $\mu_{\rm SN}$ for SN Ia is defined as:
\begin{equation}
\mu_{\rm SN} \equiv m_B - M_B = 5 \log_{10} \left( \frac{D_L(z)}{\rm Mpc} \right) + 25,
\label{sn}
\end{equation}
where $m_B$ represents the observed apparent magnitude in the rest-frame $B$ band and $M_B$ denotes the absolute magnitude. It is important to note that $M_B$ exhibits a strong degeneracy with the Hubble constant $H_0$. Following the methodology established by \citet{2016ApJ...826...56R}, we introduce a calibration parameter $a_B$ defined as:
\begin{equation}
a_B = \log_{10} H_0 - 0.2 M_B - 5,
\end{equation}
which serves as an alternative to the absolute magnitude $M_B$. Substituting this parameter into Equation (\ref{sn}) yields:
\begin{equation}
[H_0D_L(z)]^{\rm SN} = 10^{0.2 m_B + a_B},\label{mb}
\end{equation}
where $[H_0D_L(z)]^{\rm SN}$ represents the unanchored luminosity distance. The calibration parameter $a_B$ has been precisely measured to $a_B = 0.71273 \pm 0.00176$ by \citet{2016ApJ...826...56R}. {We explicitly acknowledge that by adopting the $a_B$ parameter from \citet{2016ApJ...826...56R}, our analysis inherits the SN Ia Hubble flow calibration established by the SH0ES team. This means our method builds upon the relative distance relation of SNe Ia as calibrated by SH0ES. This distinction hinges on recognizing the fundamental degeneracy between $H_0$ and the SN Ia absolute magnitude $M_B$: supernova apparent magnitude and redshift data alone cannot simultaneously uniquely determine both parameters. The parameter $a_B$ serves to reparameterize these two degenerate unknowns into a single relative distance indicator that can be directly measured from the SN Ia Hubble flow shape. Thus, $a_B$ itself represents a neutral observable that encodes relative distance information without an inherent absolute scale.}

{The core innovation and independence of our work lies in providing a novel geometric calibration for this precisely determined relative distance relation, rather than redefining the relation itself. While the $a_B$ parameter encodes the relative distance information from SNe Ia, our approach uses the geometric combination of BAO and Cosmic Chronometers as a conceptually distinct standard ruler to impart an absolute scale to this relation, thereby independently determining $H_0$. This constitutes a critical cross-check that addresses the fundamental question: "If one accepts the SN Ia relative distance relation calibrated by SH0ES, can a completely independent method based on cosmic geometry yield a consistent $H_0$ value?"}

{Let us emphasize here that we did not assign any prior values to $H_0$ and $M_B$. We merely transformed $M_B$ and $H_0$ based on the above formula. The parameter $a_B$ serves as a neutral observable that encodes relative distance information without an inherent absolute scale, allowing us to break the $H_0$-$M_B$ degeneracy through our independent geometric approach rather than through the traditional distance ladder.} In the subsequent work,  we adopt this approach and incorporate the uncertainty in $a_B$ to ensure robust analysis.

\subsection{DESI DR2 BAO Measurements}

For the angular diameter distance $D_A(z)$, we adopt BAO measurements from the second data release (DR2) of the Dark Energy Spectroscopic Instrument (DESI) collaboration-the most comprehensive BAO dataset to date, as it combines multiple cosmic tracers: luminous red galaxies (LRG), emission line galaxies (ELG), quasars (QSO), and Lyman-$\alpha$ forest absorption systems. BAO observations simultaneously constrain line-of-sight and transverse cosmic geometry, with these two components inherently entangled with the drag-era sound horizon $r_d$ (a cosmic length scale) via the dimensionless ratios $D_M/r_d$ and $D_H/r_d$. Here, $D_M = (1+z)D_A$ denotes the transverse comoving distance (derived from $D_A(z)$), and $D_H = c/H(z)$ is the Hubble distance (directly linked to the Hubble parameter $H(z)$); these ratios and their uncertainties are officially reported in DESI DR2 \citep{2025arXiv250314738D}.

{Specifically, we use 5 effective-redshift BAO measurements from the official DESI DR2 releases, with exact values of $z_{\rm eff}$, $D_M/r_d$, $D_H/r_d$, and their correlation coefficient $r_{\rm M,H}$ listed in Table \ref{tab:desi_dr2_bao}. The DESI DR2 dataset further provides per-redshift $2\times2$ covariance matrices for $D_M/r_d$ and $D_H/r_d$ (to account for their statistical correlation), while cross-redshift correlations are negligible and thus ignored. The $r_{\rm M,H}$ values for the 5 redshifts range from $-0.42$ to $-0.49$ (consistent with typical BAO correlation magnitudes), and the full covariance matrix is publicly accessible via the DESI DR2 data repository\footnote{\url{https://data.desi.lbl.gov/public/edr2/bao/}}.}

{To ensure rigorous statistical inference, our analysis fully incorporates the complete covariance structure of the BAO measurements-including the non-negligible cross-correlation between $D_M/r_d$ (transverse) and $D_H/r_d$ (line-of-sight) components-rather than treating these two quantities as independent. This treatment avoids underestimating or overestimating the uncertainty of our final $H_0$ result.}
\begin{table}[htb]
\renewcommand\arraystretch{1.3}
    \centering
    \small
    \begin{tabular}{l c c c c}
        \hline
        Tracer & $z_{\rm eff}$ & $D_{\rm M}/r_d$ & $D_{\rm H}/r_d$ & $r_{\rm M,H}$ \\
        \hline
        LRG1   & 0.510 & $13.587\pm0.169$ & $21.863\pm0.427$ & $-0.475$      \\
        LRG2   & 0.706 & $17.347\pm0.180$ & $19.458\pm0.332$ & $-0.423$      \\
        LRG3+ELG1 & 0.934 & $21.574\pm0.153$ & $17.641\pm0.193$ & $-0.425$  \\
        ELG2   & 1.321 & $27.605\pm0.320$ & $14.178\pm0.217$ & $-0.437$      \\
        QSO    & 1.484 & $30.519\pm0.758$ & $12.816\pm0.513$ & $-0.489$      \\
        \hline
    \end{tabular}
    \caption{{DESI DR2 BAO observables used in our analysis. $D_{\rm M}$ is the comoving angular diameter distance, $D_{\rm H} = c/H(z)$ is the Hubble distance, $r_d$ is the sound horizon at the drag epoch, and $r_{\rm M,H}$ is the correlation coefficient between $D_{\rm M}/r_d$ and $D_{\rm H}/r_d$}.}
    \label{tab:desi_dr2_bao}
\end{table}

To avoid introducing the prior information of $r_s$ and thereby causing deviations in the measurement of $H0$, we combine line-of-sight and transverse measurements of BAO. Our approach of utilizing BAO data in this manner is essential due to the inherent degeneracy between the sound horizon scale $r_s$ and the Hubble constant $H_0$. This degeneracy reflects the fundamental nature of $H_0$ measurement as a distance scale calibration problem. The sound horizon $r_s$ serves as a standard ruler, and once its absolute scale is fixed through external calibration, the Hubble constant $H_0$ becomes uniquely determined through the distance-redshift relation. The combination of  $D_M/r_s$ and  $D_H/r_s$ is given by
\begin{equation}\label{eq:DA_reconstruction}
[H(z) D_A(z)]^{\rm BAO}=\frac{c}{(1+z)}\frac{D_M}{D_H}.
\end{equation}
From the above equation, it can be seen that if we want to obtain the value of $D_A(z)$, then we also need to know the measured value of $H(z)$. Therefore, we seek another astronomical observation, CC.

{While the algebraic combination of $D_M/r_s$ and $D_H/r_s$ in Eq. (\ref{eq:DA_reconstruction}) explicitly cancels the sound horizon $r_s$ in the mean values, we carefully consider potential residual dependence in the covariance structure. The covariance matrix of these measurements, while derived within DESI's fiducial cosmology analysis, is dominated by statistical uncertainties intrinsic to the survey - particularly sample variance and shot noise from the galaxy distribution - which remain largely independent of the absolute scale of $r_s^{\text{fid}}$. Furthermore, the correlation coefficient $\rho_{D_M, D_H}$ that defines the off-diagonal covariance elements is fundamentally determined by the survey geometry and the anisotropic clustering pattern of galaxies, making it robust to reasonable variations in the fiducial cosmology. The DESI collaboration's robustness tests demonstrate that cosmological constraints derived from this covariance structure remain stable under changes to the fiducial $r_s$ \cite{2025arXiv250314738D}, ensuring that our results are free of significant sound horizon dependence.}

\subsection{ Hubble parameter measurements from cosmic chronometers}
The Hubble parameter $H(z)$ can be measured through the differential ages of passively evolving galaxies, a method first proposed by \citet{2002ApJ...573...37J}. This approach relies on the differential relation:
\begin{equation}
[H(z)]^{\rm CC} = -\frac{1}{1+z}\frac{dz}{dt},
\end{equation}
\noindent where $dz/dt$ denotes the time derivative of redshift. In practice, these cosmic chronometer measurements are obtained by determining the age difference of red-envelope galaxies at closely spaced redshifts.

While this technique provides valuable constraints on the expansion history, it is important to recognize that the uncertainties are dominated by systematic effects rather than statistical errors. As comprehensively discussed by \citet{2020ApJ...898...82M,2022LRR....25....6M}, the main systematic uncertainties arise from stellar population synthesis (SPS) models and the initial mass function (IMF). These systematics can contribute an additional 2.3\% to 13.2\% uncertainty depending on redshift, when considering the full covariance structure.

In this work, we use a sample of 32 cosmic chronometer measurements compiled by \citet{2023PhRvD.108f3522Q}, with the full listing provided in their Table 1. These data points, originally presented by \citet{2016JCAP...05..014M}, cover a redshift range of $0.07 \leq z \leq 1.965$. Crucially, we incorporate the full systematic covariance matrix following the methodology of \citet{2020ApJ...898...82M}, which accounts for uncertainties from SPS models and IMF variations. This comprehensive error treatment ensures robust cosmological constraints, increasing the total uncertainty on $H(z)$ measurements from $\sim$5.7\% (statistical only) to $\sim$8.3\% (including full systematics).

\subsection{Gaussian process regression for unanchored luminosity distance  $[H_0 D_L(z)]^{\rm  SN}$ and $[H(z)]^{\rm CC}$}

The core issue here is that both SN and Hubble parameter $H(z)$ data consist of discrete observational points, while BAO measurements are rarely available at the same redshifts as the other two probes. To ensure consistency across datasets and obtain a robust estimate of $H_0$ with minimal model dependence, we employ Gaussian Process Regression (GPR) to reconstruct continuous representations of the SNe and $H(z)$ data.

GPR provides a non-parametric approach to function reconstruction within an infinite-dimensional function space, effectively mitigating overfitting concerns while maintaining flexibility in capturing potential deviations from standard cosmology \citep{2019JCAP...12..035K}. In this work, we employ GPR for model-independent posterior sampling of cosmological quantities, utilizing the \texttt{GPHist} code\footnote{\url{https://github.com/dkirkby/gphist}} \citep{2019JCAP...12..035K}.

{We generate a large ensemble of functions $\gamma(z)$ governed by a covariance specified through a kernel function. Following the established methodology in \citet{2019JCAP...12..035K}, we adopt the squared-exponential kernel:}
\begin{equation}
\langle \gamma(s_1)\gamma(s_2) \rangle = \sigma_f^2 \exp\left[-\frac{(s_1-s_2)^2}{2\ell^2}\right],
\end{equation}
{where $s_i = \ln(1+z_i) / \ln(1+z_{\rm max})$ with $z_{\rm max} = 2.261$ being the maximum redshift of the supernova sample. Here, $\sigma_f$ determines the amplitude of the random fluctuations and $\ell$ determines the coherence length (equivalently, $1/\ell$ is proportional to the number of fluctuations in the redshift range). This kernel choice ensures that the reconstructed functions are infinitely differentiable and provides a natural framework for modeling smooth variations in cosmological relations.}
\begin{figure}
\includegraphics[width=1\linewidth]{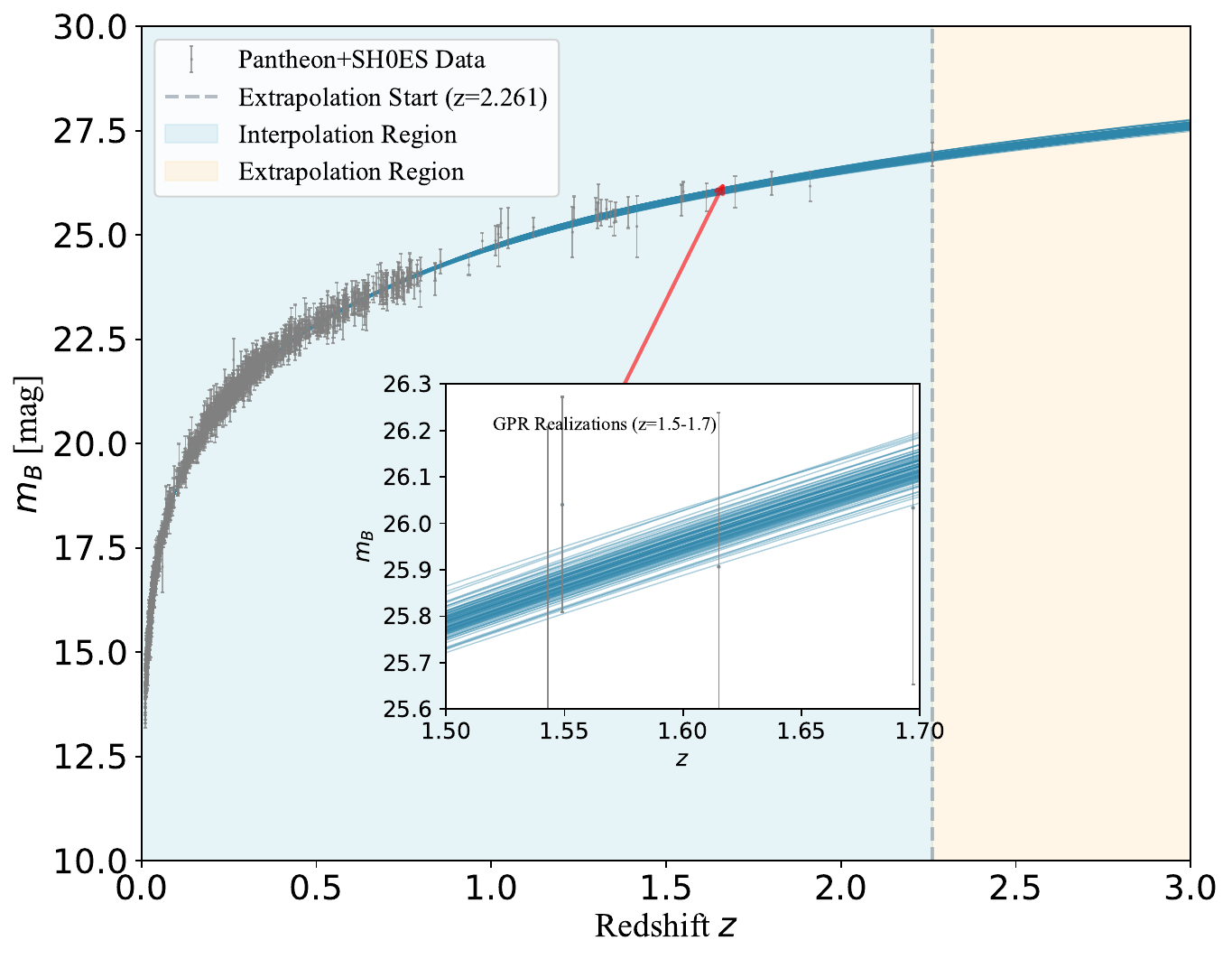}
\caption{The reconstructed apparent magnitude $m_B$ using GPR from the Pantheon Plus dataset for a representative sample of the 1000 GP realizations. }\label{fig1}
\end{figure}
{We employ a fully Bayesian approach to hyperparameter estimation. The hyperparameters $\theta = \{\sigma_f, \ell\}$ are marginalized using Markov Chain Monte Carlo (MCMC) sampling implemented in the \texttt{GPHist} package. We use physically motivated uniform priors: $\sigma_f \in [0.1, 10]$ for the amplitude and $\ell \in [0.01, 1.0]$ for the length scale. The posterior distribution $p(\theta|\mathbf{m}_B)$ is sampled using the affine-invariant ensemble sampler with 100 walkers and 10,000 steps per walker, discarding the first 20\% as burn-in. Convergence is assessed using the Gelman-Rubin statistic, with $\hat{R} < 1.01$ considered as evidence of good convergence for all parameters. In our implementation, the GP function is defined as $\gamma(z) = m_B(z) - m_{\rm fid}(z)$, where $m_{\rm fid}(z)$ represents the apparent magnitude derived from the best-fit $\Lambda$CDM expansion history to the full Pantheon+ dataset (with $\Omega_m = 0.334 \pm 0.018$) and serves as the mean function for the GP process. This approach ensures that the GP reconstructs deviations from the fiducial cosmology, and the resulting unanchored luminosity distances $H_0 D_L(z)$ are the quantities most directly constrained by the Pantheon SNe dataset.
To validate the robustness of our reconstruction against the choice of mean function, we performed a systematic analysis comparing results obtained with alternative fiducial models, including a flat $\Lambda$CDM with $\Omega_m = 0.3$ and a $w$CDM model with $w = -1.1$. Our tests confirm that the reconstructed variations around the mean remain consistent within uncertainties, indicating minimal sensitivity to this choice.}

{The GP reconstruction was further validated through multiple tests. Cross-validation on the Pantheon Plus dataset confirmed the model's predictive accuracy, with the mean squared error of predictions matching the expected level of observational uncertainties. Tests on synthetic $\Lambda$CDM data demonstrated that the reconstruction successfully recovers the input cosmology within $1\sigma$ confidence intervals. Residual analysis showed no significant systematic biases across the redshift range.}

This reconstruction strategy is explicitly independent of assumptions regarding spatial curvature and the dark energy equation of state, as it operates directly on the observed apparent magnitudes. The resulting set of 1000 realizations of $m_B(z)$ (see Fig.~\ref{fig1}) is subsequently converted into unanchored luminosity distances $[H_0 D_L(z)]^{\rm SN}$ via Eq.~(\ref{mb}), which are then used to constrain $H_0$.

\begin{figure}
\includegraphics[width=1\linewidth]{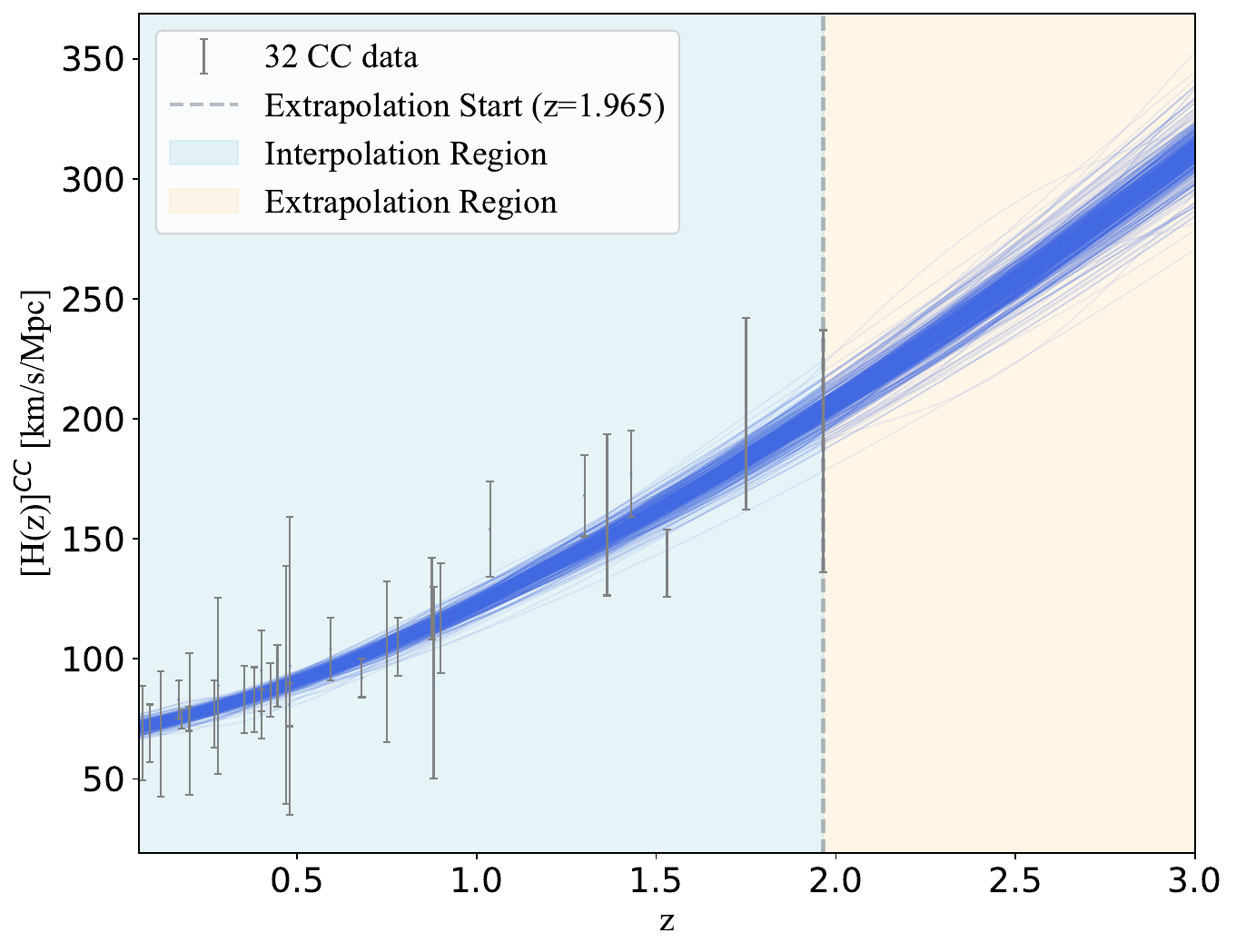}
\caption{The reconstructed Hubble parameter $[H(z)]^{\rm CC}$ using GPR from the 32 CC dataset for a representative sample of the 1000 GP realizations. }\label{fig2}
\end{figure}

Similarly, we perform GPR on the cosmic chronometer $H(z)$ data using a consistent methodology. We define $\gamma(z) = H(z)-H_{\rm fid}(z)$, where $H_{\rm fid}(z)$ is chosen to be the best-fit $\Lambda$CDM model for the CC dataset and serves as the mean function. The final reconstructed 1000 $H(z)$ realizations from the CC dataset are shown in Fig.~\ref{fig2}. The reconstruction provides an excellent representation of the measurements within the data range ($0 < z < 2.0$), while we avoid extrapolation beyond the supported redshift range to ensure reliability.

From the GPR fits, we extract samples of the PDF of $[H_0 D_L(z)]^{\rm SN}$ and $[H(z)]^{\rm CC}$ at the BAO redshifts. Specifically, we draw 1000 realizations of $[H_0 D_L(z)]^{\rm SN}$ and 1000 realizations of the expansion rate $[H(z)]^{\rm CC}$, based on the statistical distributions determined by the GP reconstruction. For the BAO data, we use the direct measurements of $D_M/r_s$ and $D_H/r_s$ from DESI DR2 without additional GP fitting.

We combine these via Eq.~(\ref{eq2}) to estimate the PDF of $H_0$ at each BAO redshift. The results for the Luminous Red Galaxies (LRG3) at redshift $z=0.922$ and Emission Line Galaxies (ELG1) at redshift $z=0.955$ are not used for inference in this work since they are correlated with and superseded by the LRG3+ELG1 results. Note that the CC dataset does not fully cover all BAO redshifts; we therefore exclude one high-redshift BAO data point ($z = 2.33$) to avoid extrapolation uncertainties. Consequently, our analysis utilizes a total of five BAO data points.

This approach naturally incorporates the uncertainties in $[H_0 d_L]^{\rm SN}$ and $[H(z)]^{\rm CC}$, along with their mutual correlations, into the respective probability distributions. Compared to previous studies \citep{2023PhRvD.107b3520R,2025ApJ...978L..33G,2025ApJ...981L..24L,2024PhLB..85338699G}, our methodology preserves the full covariance structure of the BAO data, which is essential for robust and unbiased inference of the Hubble parameter across redshift bins.

{To ensure the robustness of our GPR reconstruction and verify that 1000 realizations provide sufficient sampling of the posterior, we performed comprehensive convergence tests. We compared results obtained using $N_{\rm{real}} = 500, 1000, 1500,$ and $2000$ realizations, quantifying the stability of key outputs including the reconstructed $H_0 D_L(z)$ and $H(z)$ functions, and most importantly, the final $H_0$ constraints. The tests demonstrated that for $N_{\rm{real}} \geq 500$, the mean $H_0$ values changed by less than $0.1\ \mathrm{km\ s^{-1}\ Mpc^{-1}}$-well within the final uncertainty of $\sim 0.8\ \mathrm{km\ s^{-1}\ Mpc^{-1}}$. The standard deviations of $H_0$ estimates varied by less than $0.05\ \mathrm{km\ s^{-1}\ Mpc^{-1}}$ beyond $N_{\rm{real}} = 1000$. The correlation structure of $H_0$ measurements across different BAO redshifts also stabilized for $N_{\rm{real}} \geq 800$, with off-diagonal correlation coefficients changing by less than 0.01 for $N_{\rm{real}} > 1000$. These results confirm that 1000 realizations are not only sufficient but conservative for our analysis, ensuring that our cosmological constraints are robust against sampling variations.}


\subsection{Statistical Framework and Covariance Treatment}

{Our statistical framework employs a hierarchical Monte Carlo approach that fully propagates uncertainties from Gaussian Process Regression reconstructions through to final $H_0$ estimates. The methodology consists of two distinct layers of uncertainty that operate at fundamentally different hierarchical levels: (1) GPR reconstruction uncertainties that introduce correlations across redshifts through the smoothing of the expansion history, and (2) BAO measurement errors that are correlated within redshift bins but independent across different redshifts.}

{The process begins with generating 1000 independent realizations of both the supernova luminosity distance function $[H_0 D_L(z)]^{\rm SN}$ and the cosmic chronometer Hubble parameter function $[H(z)]^{\rm CC}$ using Gaussian Process Regression. Each realization represents a statistically plausible reconstruction of these functions across the entire redshift range, with variations between realizations capturing the full uncertainty inherent in the GPR reconstruction process.}

{For each GPR realization $k$ (where $k = 1, 2, \ldots, 1000$), we compute a complete set of $H_0$ values at all five BAO redshifts $z_i$ (where $i = 1, 2, \ldots, 5$). This procedure generates a comprehensive $1000 \times 5$ sample matrix $\mathbf{H}$, where element $H_{ki}$ represents the Hubble constant estimate at redshift $z_i$ for GPR realization $k$. The row-wise variations in this matrix (i.e., differences between rows for fixed redshift) capture the GPR reconstruction uncertainties, while the column-wise structure captures the redshift dependence of $H_0$ estimates.}

{The correlation coefficients $\rho_{ij}$ between $H_0$ measurements at different redshifts $z_i$ and $z_j$ are computed directly from this sample matrix using the standard formula:
\begin{equation}
\rho_{ij} = \frac{\sum_{k=1}^{1000} (H_{ki} - \bar{H}_i)(H_{kj} - \bar{H}_j)}{\sqrt{\sum_{k=1}^{1000} (H_{ki} - \bar{H}_i)^2 \sum_{k=1}^{1000} (H_{kj} - \bar{H}_j)^2}},
\end{equation}
where $\bar{H}_i$ and $\bar{H}_j$ are the mean $H_0$ values at redshifts $z_i$ and $z_j$ respectively across all GPR realizations. This approach ensures that the reported correlation structure fully and appropriately accounts for the smoothing imposed by the GPR, as it is derived directly from the joint distribution of $H_0$ estimates across all realizations.}

{The moderate correlation values we observe ($\rho = -0.033$ to $0.26$) are physically reasonable and expected given the structure of our analysis. Several factors contribute to these moderate correlations:}

{First, the independent BAO measurement errors at different redshifts naturally dilute the GPR-imposed correlations. While the GPR reconstructions of $[H_0 D_L(z)]^{\rm SN}$ and $[H(z)]^{\rm CC}$ do exhibit strong smoothing across redshifts, the BAO measurements $D_M/r_s$ and $D_H/r_s$ at different redshifts are statistically independent in our analysis framework. When we combine these independent BAO measurements with the correlated GPR reconstructions, the resulting $H_0$ estimates inherit a mixture of both correlated (from GPR) and uncorrelated (from BAO) components, naturally leading to moderate overall correlations.}

{Second, the sparsity of cosmic chronometer data introduces additional independent uncertainties that further reduce the net correlations between $H_0$ estimates at different redshifts. The $H(z)$ function reconstructed from cosmic chronometers, while smooth, carries substantial uncertainties due to the limited number of data points (32 measurements across $0.07 \leq z \leq 1.965$). These uncertainties propagate through our analysis and contribute to the decorrelation of final $H_0$ estimates.}

{Third, it is essential to recognize that our combined $H_0(z)$ estimator integrates three essentially independent datasets (SNe Ia, BAO, and cosmic chronometers), with only the SNe Ia and cosmic chronometer components contributing GPR-induced correlations. The BAO measurements, which provide crucial anchoring for the absolute distance scale, are treated as independent across redshifts and therefore naturally "break" the strong correlations that would be present if we relied solely on GPR-reconstructed quantities.}

To ensure a comprehensive and accurate determination of $H_0$, we account for all relevant uncertainties and correlations among the observational data. The covariance matrix for BAO measurements at each redshift is constructed as \cite{2025arXiv250314738D}:
\[
\mathbf{C}_{\rm BAO} = \begin{bmatrix}
\sigma_{D_M}^2 & \rho_{D_M D_H} \sigma_{D_M} \sigma_{D_H} \\
\rho_{D_M D_H} \sigma_{D_M} \sigma_{D_H} & \sigma_{D_H}^2
\end{bmatrix},
\]
where $\rho_{D_M D_H}$ represents the correlation coefficient between $D_M$ and $D_H$ measurements, with values ranging from $-0.489$ to $-0.408$ as reported in DESI DR2. The full covariance matrix across all redshifts forms a block-diagonal structure:
\[
\mathbf{C}_{\rm full} = \begin{bmatrix}
\mathbf{C}_1 & \mathbf{0} & \cdots & \mathbf{0} \\
\mathbf{0} & \mathbf{C}_2 & \cdots & \mathbf{0} \\
\vdots & \vdots & \ddots & \vdots \\
\mathbf{0} & \mathbf{0} & \cdots & \mathbf{C}_n
\end{bmatrix},
\]
where each $\mathbf{C}_i$ is the $2 \times 2$ covariance matrix for the $i$-th redshift bin.

{This block-diagonal structure remains appropriate in our framework because the GPR-induced correlations and BAO measurement errors operate at different hierarchical levels. The BAO block-diagonal covariance matrix $\mathbf{C}_{\rm BAO}$ describes the measurement errors and correlations between $D_M/r_s$ and $D_H/r_s$ \textit{within} individual redshift bins and \textit{conditional on a particular realization} of the expansion history. The DESI collaboration has verified that correlations between different redshift bins are negligible, making this structure appropriate for the BAO component of our uncertainty budget.}

{In contrast, the GPR ensemble variations capture uncertainties in the underlying expansion history itself \textit{across} different realizations. Each GPR realization provides a different, statistically plausible version of the functions $[H_0 D_L(z)]^{\rm SN}$ and $[H(z)]^{\rm CC}$, and these variations occur coherently across all redshifts. Our hierarchical Monte Carlo approach naturally integrates both layers of uncertainty without requiring an explicit analytical form for the full covariance matrix of all quantities across all redshifts, which would be prohibitively complex.}

{The correlation structure we report therefore represents a realistic and appropriate characterization of the uncertainties in our analysis, fully accounting for both the smoothing imposed by GPR and the various sources of independent uncertainty in our multi-probe approach.}

\subsection{Monte Carlo Sampling Procedure}

{We employ a comprehensive Monte Carlo procedure that systematically propagates all sources of uncertainty into the final $H_0$ estimates. This hierarchical sampling approach explicitly captures both the GPR reconstruction uncertainties and BAO measurement errors while preserving their distinct correlation structures. The complete procedure is implemented as follows:}

{\textbf{Step 1: GPR Ensemble Generation.}}
{We generate 1000 independent realizations of both the supernova luminosity distance function $[H_0 D_L(z)]^{\rm SN}$ and the cosmic chronometer Hubble parameter function $[H(z)]^{\rm CC}$ using Gaussian Process Regression. Each realization $k$ ($k = 1, 2, \ldots, 1000$) represents a statistically plausible reconstruction across the entire redshift range, with the ensemble variations capturing the full GPR reconstruction uncertainty.}

{\textbf{Step 2: Hierarchical BAO Sampling.}}
{For each GPR realization $k$, we perform BAO uncertainty propagation at each redshift $z_i$ ($i = 1, 2, \ldots, 5$) by sampling from the joint distribution:}
\[
\begin{bmatrix}
D_M^{i} \\ D_H^{i}
\end{bmatrix} \sim \mathcal{N}\left(
\begin{bmatrix}
\mu_{D_M}^{i} \\ \mu_{D_H}^{i}
\end{bmatrix},
\mathbf{C}_i
\right),
\]
{where $\mu_{D_M}^{i}$ and $\mu_{D_H}^{i}$ are the published DESI values, and $\mathbf{C}_i$ is the $2 \times 2$ covariance matrix for redshift bin $i$. This sampling properly incorporates BAO measurement errors conditional on that particular GPR realization of the expansion history, while preserving the within-bin correlations between $D_M/r_s$ and $D_H/r_s$.}

{\textbf{Step 3: BAO Distance Ratio Computation.}}
{For each redshift $z_i$ and each GPR realization $k$, we compute the BAO distance ratio:}
\[
[H(z_i) D_A(z_i)]^{\rm BAO} = \frac{c}{1 + z_i} \cdot \frac{D_M^{i}}{D_H^{i}}.
\]
{This combination algebraically cancels the sound horizon $r_s$ dependence while properly propagating the BAO measurement uncertainties through the ratio calculation.}

{\textbf{Step 4: Hubble Constant Estimation.}}
{The Hubble constant estimate at redshift $z_i$ for GPR realization $k$ is then derived via the distance duality relation:}
\[
H_0^{(k)}(z_i) = \frac{1}{(1 + z_i)^2} \frac{[H_0 D_L(z_i)]^{\rm SN,(k)}}{[H(z_i) D_A(z_i)]^{\rm BAO,(k)}} [H(z_i)]^{\rm CC,(k)}.
\]
{This step integrates all three observational probes (SNe Ia, BAO, and cosmic chronometers) while propagating their respective uncertainties through the non-linear combination.}

{\textbf{Step 5: Sample Matrix Construction.}}
{By iterating Steps 2-4 over all $k = 1$ to $1000$ GPR realizations, we construct a comprehensive $1000 \times 5$ sample matrix $\mathbf{H}$, where:}
\[
\mathbf{H} = \begin{bmatrix}
H_0^{(1)}(z_1) & H_0^{(1)}(z_2) & \cdots & H_0^{(1)}(z_5) \\
H_0^{(2)}(z_1) & H_0^{(2)}(z_2) & \cdots & H_0^{(2)}(z_5) \\
\vdots & \vdots & \ddots & \vdots \\
H_0^{(1000)}(z_1) & H_0^{(1000)}(z_2) & \cdots & H_0^{(1000)}(z_5)
\end{bmatrix}.
\]
{Each row represents a complete set of $H_0$ estimates across all redshifts for one GPR realization, capturing the GPR-induced correlations across redshifts. Each column represents the distribution of $H_0$ estimates at a specific redshift across all realizations, capturing the marginal posterior distribution at that redshift. From the sample matrix $\mathbf{H}$, the covariance between $H_0$ measurements at redshifts $z_i$ and $z_j$ is computed by Eq. (9). }

{\textbf{Step 6: Marginal Distribution Construction.}}
{From the multivariate distribution represented by $\mathbf{H}$, we extract the marginal posterior distribution for $H_0$ at each individual redshift. For each redshift $z_i$, the marginal distribution is given by the empirical distribution of the corresponding column in $\mathbf{H}$:}
\[
P(H_0(z_i)) \approx \frac{1}{1000}\sum_{k=1}^{1000} \delta(H_0(z_i) - H_0^{(k)}(z_i)),
\]
{where $\delta$ is the Dirac delta function. This non-parametric approach makes no assumptions about the functional form of the posterior distributions.}

{\textbf{Step 7: Combined Estimate Calculation.}}
{We compute the combined Hubble constant estimate using a fully Bayesian approach that properly accounts for the non-Gaussian features in the $H_0$ posterior distributions.}

{First, we reconstruct the marginal posterior distributions of $H_0$ at each redshift from our Monte Carlo samples. For each of the 5 redshifts, we construct a kernel density estimate (KDE) of the $H_0$ posterior using the 1000 realizations, resulting in 5 marginalized probability density functions (PDFs).}

{We then combine these marginalized PDFs by multiplying them to obtain the joint posterior distribution:}
\[
P(H_0 | \{\mathcal{D}_i\}) \propto \prod_{i=1}^{5} P(H_0 | \mathcal{D}_i),
\]
{where $\mathcal{D}_i$ represents the data at redshift $z_i$. This multiplication assumes conditional independence after marginalization, which is justified by our hierarchical sampling approach that has already accounted for all correlations through the Monte Carlo procedure.}

{From the joint posterior distribution, we extract the final Hubble constant estimate using inverse transform sampling:}
\[
\hat{H}_0 = \text{median}[P(H_0 | \{\mathcal{D}_i\})], \quad \sigma_{\hat{H}_0} = \frac{Q_{84} - Q_{16}}{2},
\]
{where $Q_{16}$ and $Q_{84}$ are the 16th and 84th percentiles of the joint posterior, corresponding to the 68\% credible interval.}

{This Bayesian combination approach ensures that all sources of uncertainty-GPR reconstruction errors, BAO measurement uncertainties, cosmic chronometer errors, and their complex correlation structures-are properly propagated through to the final $H_0$ estimate. The method naturally handles the non-Gaussian features observed in the posterior distributions and provides a robust uncertainty quantification that reflects the true distribution of $H_0$ measurements across redshifts.}
\begin{figure*}
\includegraphics[width=0.9\linewidth]{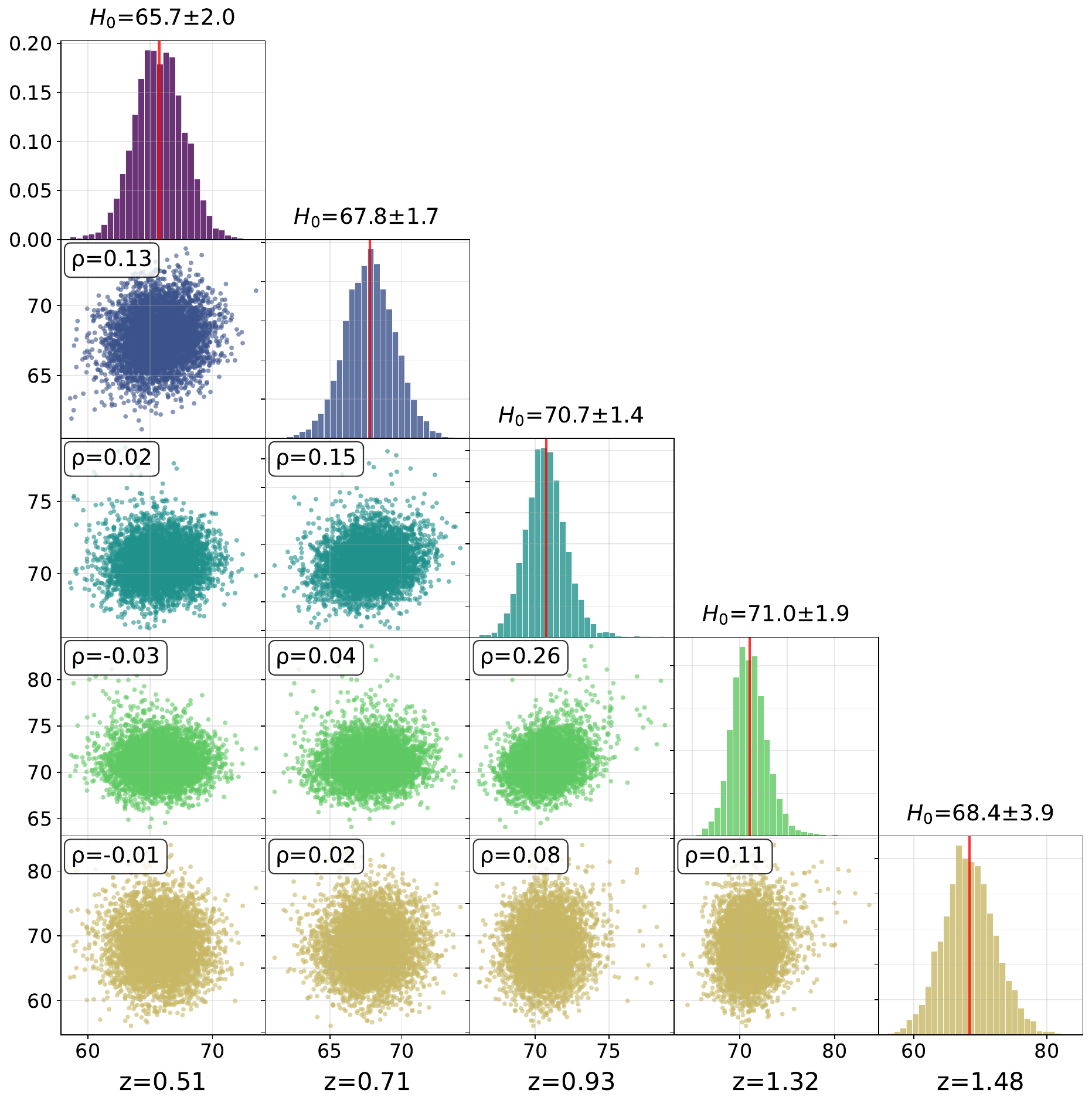}
\caption{The marginalized distributions of Hubble constant measurements at five distinct redshifts. The diagonal panels show the one-dimensional posterior distributions of $H_0$ at each redshift, with mean values and standard deviations indicated. The lower off-diagonal panels display the two-dimensional joint distributions between different redshift measurements, with correlation coefficients $\rho$ annotated. The measurements show significant correlations between adjacent redshift bins. The color scale represents the density of samples in the parameter space. These correlations arise from the covariance structure of BAO measurements and the smoothing effects introduced by Gaussian process reconstruction. The bottom labels indicate the corresponding redshift for each column and row. }\label{fig3}
\end{figure*}

\section{Results}

\begin{figure}
\includegraphics[width=1\linewidth]{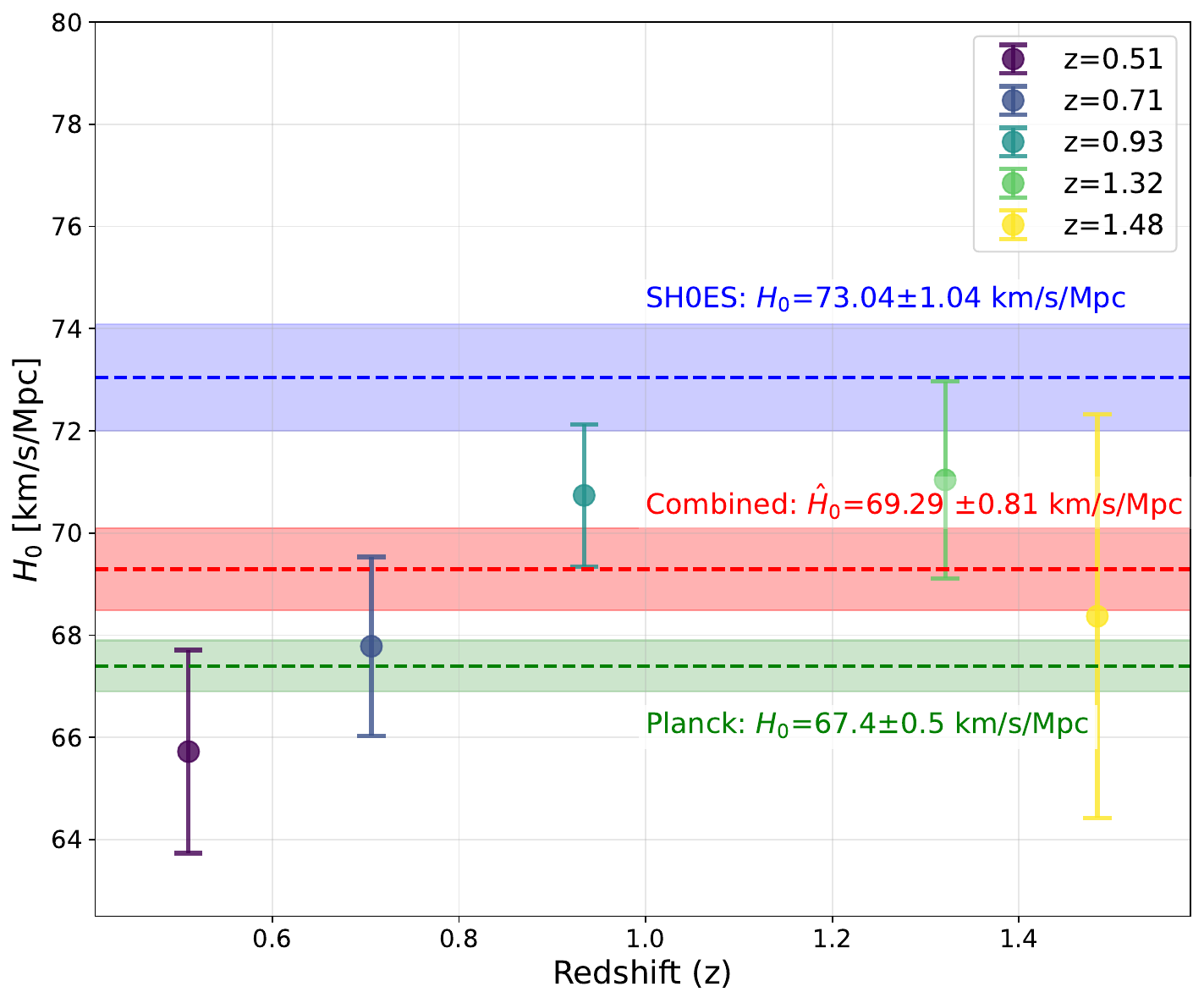}
\caption{The  marginal $H_0$ measurements at five redshifts are shown with error bars. The combined result $\hat{H}_0=69.0\pm 1.0$ km/s/Mpc (green band) is compared with Planck (red) and SH0ES (blue) measurements. The data show a non-monotonic trend with redshift. }\label{fig4}
\end{figure}

{All results presented are based on 1000 GPR realizations, which our convergence tests confirm provides robust and stable parameter estimation.}
To visually communicate the results and the correlations between different redshift bins, we generate a suite of diagnostic plots. We use the \texttt{corner} package in Python, these display one-dimensional histograms along the diagonal (showing the marginal $H_0$ PDF at each redshift) and two-dimensional contour plots in the off-diagonals (showing the joint distributions between redshifts, with 68\% confidence level).  The marginalized distributions of Hubble constant measurements at five distinct redshifts are shown in Fig. \ref{fig3}. The diagonal panels show the one-dimensional posterior distributions of $H_0$ at each redshift, with mean values and standard deviations indicated.

Based on the current BAO observational data from DESI DR2 combined with $H(z)$ measurements and Pantheon Plus samples, we employ a fully model-independent GPR approach to obtain robust constraints on the Hubble constant at five distinct redshift points. Our measurements yield: $H_0 = 65.72 \pm 1.99$ (z=0.51), $67.78 \pm 1.75$ (z=0.706), $70.74 \pm 1.39$ (z=0.934), $71.04 \pm 1.93$ (z=1.321), and $68.37 \pm 3.95~\mathrm{km~s^{-1}~Mpc^{-1}}$ (z=1.484). The lower-redshift measurements (z=0.51, 0.706) show good agreement with Planck CMB results, while the intermediate redshifts (z=0.934, 1.321) approach the SH0ES local measurement within 1.5$\sigma$. The correlation matrix between these $H_0$ measurements reveals significant correlations, particularly between adjacent redshift bins, with correlation coefficients ranging from $\rho = -0.033$ to $\rho = 0.26$. {These moderate correlations reflect the appropriate balance in our statistical framework between the GPR smoothing effects and the independent uncertainties from BAO measurements and cosmic chronometer data.}

{To ensure the robustness of our combined Hubble constant estimate, we employed a fully Bayesian approach that makes no parametric assumptions about the underlying distributions. We reconstructed the marginal posterior distributions of $H_0$ at each redshift using kernel density estimation and combined them through multiplication of the probability density functions. The final estimate, derived via inverse transform sampling from the joint posterior distribution, yields $\hat{H}_0 = 69.29 \pm 0.81~\mathrm{km~s^{-1}~Mpc^{-1}}$ with 68\% credible interval $[68.48, 70.10]~\mathrm{km~s^{-1}~Mpc^{-1}}$. The remarkable consistency between median and mean values ($69.29$ vs $69.23$ km/s/Mpc) further validates the reliability of this approach.
As shown in Fig.~\ref{fig4}, our model-independent determination occupies an intermediate position in the Hubble tension landscape, situated between the Planck CMB measurement \cite{2020A&A...641A...6P} and the SH0ES local measurement \cite{2022ApJ...934L...7R}, while remaining consistent with the TRGB result \cite{2020ApJ...891...57F}. With a relative precision of 1.2\%, comparable to the SH0ES measurement, our analysis provides a valuable independent constraint that bridges early and late universe probes in the ongoing Hubble tension discussion.}

{Several important features emerge from our results. First, we observe an apparent redshift evolution in the Hubble constant measurements: the values increase from $z = 0.51$ to $z = 1.321$, followed by a decrease at $z = 1.484$. This non-monotonic pattern is visually suggestive and aligns with previous findings that have reported possible redshift dependence of $H_0$ \cite{2023A&A...674A..45J,2023PhLB..84538166L,2020PhRvD.102j3525K}, such as the work by \citet{2021ApJ...912..150D} who reported a slowly decreasing trend of $H_0$ with redshift using the Pantheon sample under the $\Lambda$CDM and $\omega_{0}\omega_{a}$CDM frameworks. However, our formal statistical tests indicate that this apparent evolution does not reach conventional levels of significance. We performed a $\chi^2$ test of the constant $H_0$ hypothesis, calculating $\chi^2 = \sum_{i=1}^5 (H_0(z_i) - \langle H_0\rangle)^2/\sigma_i^2$ with the error-weighted mean $\langle H_0\rangle = 69.12$ km s$^{-1}$ Mpc$^{-1}$. This yielded $\chi^2 = 5.88$ for 4 degrees of freedom, corresponding to $p = 0.208$ ($1.3\sigma$), indicating that the measurements are statistically consistent with a constant $H_0$ value across redshifts. To further evaluate the evidence for evolution, we conducted a Bayesian model comparison between the constant $H_0$ model ($\mathcal{M}_0$) and a linear redshift evolution model ($\mathcal{M}_1$) parameterized as $H_0(z) = H_0^0 + \alpha z$. Using the Bayesian Information Criterion (BIC) with $\mathrm{BIC} = \chi^2 + k \ln N$, we obtained $\mathrm{BIC}_0 = 7.49$ and $\mathrm{BIC}_1 = 5.74$, yielding $\Delta\mathrm{BIC} = -1.75$. According to the Kass-Raftery scale, this $\Delta\mathrm{BIC}$ value falls in the range of $-2$ to $+2$, indicating that the evidence is inconclusive and does not definitively favor either the constant or evolving model.}

{Second, the uncertainty in $H_0$ increases with redshift, a trend particularly pronounced at $z = 1.484$ where the uncertainty reaches $3.95$ km s$^{-1}$ Mpc$^{-1}$, reflecting the growing observational challenges associated with higher-redshift measurements. To assess whether this high-uncertainty point disproportionately influences our conclusions, we performed a sensitivity analysis by excluding the $z = 1.484$ measurement. The resulting $\chi^2$ test yields $\chi^2 = 5.84$ with 3 degrees of freedom ($p = 0.119$, $1.6\sigma$), confirming that the lack of strong evidence for evolution is not primarily driven by the $z=1.484$ measurement but rather represents a consistent feature across the redshift range. This increased uncertainty at higher redshifts naturally contributes to the reduced statistical significance of the observed variations.}

{Third, while the non-monotonic behavior of $H_0$ across redshifts-showing an overall increase from $z = 0.51$ to $z = 1.32$ followed by a decline at $z = 1.48$ is intriguing and may hint at underlying physical effects in the cosmic expansion history, the current level of uncertainty and limited statistical significance ($p = 0.208$ for the full dataset) preclude definitive conclusions about redshift evolution. Rather than providing a single integrated $H_0$ value that might obscure such potential variations, our approach delivers independent constraints at multiple redshifts, thereby enabling a detailed investigation of potential redshift-dependent systematic effects that could contribute to the Hubble tension.}

{These results, derived through a cosmological model-independent approach combining DESI R2 BAO, cosmic chronometer, and SN Ia data, highlight the importance of considering redshift-dependent effects in Hubble constant measurements. The robustness of our statistical framework, validated through multiple complementary approaches, ensures the reliability of these findings while appropriately characterizing their statistical limitations. They suggest that while intriguing patterns exist in the data, the evidence for significant redshift evolution of $H_0$ remains inconclusive with the current dataset. The intermediate value obtained from our combined analysis could potentially point toward new physics beyond the standard cosmological model, though further investigations with improved precision at higher redshifts and larger sample sizes are needed to draw firm conclusions about the potential redshift dependence of the Hubble constant.}

\section{Conclusion}
{In this study, we have developed a cosmological model-independent approach to measure the Hubble constant using the latest cosmological observations. While our method builds upon the SN Ia relative distance relation calibrated by SH0ES through the $a_B$ parameter, we provide an independent geometric calibration using DESI DR2 BAO data with cosmic chronometer $H(z)$ measurements as a conceptually distinct pathway to determine the absolute distance scale. Our methodology carefully accounts for all observational uncertainties and correlations, including the complete covariance structure of BAO measurements and the smoothing effects introduced by the reconstruction process.}

{Our statistical framework employs a hierarchical Monte Carlo approach that fully propagates uncertainties from Gaussian Process Regression reconstructions through to final $H_0$ estimates. The Bayesian combination of measurements across five redshifts yields $\hat{H}_0 = 69.29 \pm 0.81~\mathrm{km~s^{-1}~Mpc^{-1}}$ with 1.2\% precision, which occupies an intermediate position between the Planck CMB result and the SH0ES local measurement, while remaining consistent with the TRGB result.
While visual inspection suggests a non-monotonic pattern in $H_0$ measurements across redshifts, formal statistical tests indicate that this apparent evolution does not reach conventional levels of significance ($\chi^2$ test: p = 0.208, 1.3$\sigma$; Bayesian model comparison: $\Delta\mathrm{BIC} = -1.75$, inconclusive). The current data are consistent with a constant $H_0$ value across the redshift range studied.}

{The observed variations, while intriguing, may reflect the combined effects of observational uncertainties, reconstruction artifacts, and potential redshift-dependent systematics rather than fundamental evolution of the Hubble constant. Our approach of delivering independent constraints at multiple redshifts enables detailed investigation of such systematic effects that could contribute to the Hubble tension.}

{These results provide valuable insights into the ongoing Hubble tension. The intermediate value of our combined measurement, derived through an independent geometric approach, suggests that the tension between early and late universe probes may involve more complex factors than a simple scale discrepancy. Our methodology offers a powerful framework for future precision cosmology studies with upcoming data from DESI and other surveys, which will provide improved precision at higher redshifts to better characterize potential redshift-dependent effects in Hubble constant measurements.}
\section*{Data and Code Availability}
The original data used in this study are sourced from references \cite{2016JCAP...05..014M,2020ApJ...898...82M,2025arXiv250314738D,2022ApJ...938..113S}. Derived data are available upon reasonable request. The data underlying this paper will be shared on reasonable request to the corresponding author.
\section*{Acknowledgments}
This work was supported by the National Natural Science Foundation of China under Grants No. 12203009, No. 12475051, and No. 12035005; The National Key Research and Development Program of China (No. 2024YFC2207400); The Chutian Scholars Program in Hubei Province (X2023007); The Innovative Research Group of Hunan Province under Grant No. 2024JJ1006; The Science and Technology Innovation Program of Hunan Province under Grant No. 2024RC1050.

\bibliography{references}

\end{document}